\documentclass[aps,prl,amsmath,twocolumn]{revtex4}
\usepackage{graphicx,dcolumn,bm,amssymb,amsmath}
\usepackage{color,wrapfig,braket}
\usepackage{hyperref,placeins,ulem}
\usepackage[caption=false]{subfig}
\usepackage{yhmath}

\def\ket#1{\left|#1\right\rangle}

\begin{document}

\title{Crystal-symmetry preserving Wannier states for
fractional Chern insulators}
\author{Chao-Ming Jian}
\author{Xiao-Liang Qi}
\affiliation{Department of Physics, Stanford University, Stanford, California 94305, USA.}
\date{\today}

\begin{abstract}
Recently, many numerical evidences of fractional Chern insulator, i.e. the fractional quantum Hall states on lattices, are proposed when a Chern band is partially filled. Some trial wave functions of fractional Chern insulators can be obtained by mapping the fractional quantum Hall wave functions defined in the continuum onto the lattice through the Wannier state representation (Phys. Rev. Lett. {\bf 107}, 126803 (2011)) in which the single particle Landau orbits in the Landau levels are identified with the one-dimensional Wannier states of the Chern bands with Chern number $C=1$.
However, this mapping generically breaks the lattice point group symmetry. In this paper, we discuss a general approach of modifying the mapping to accommodate the lattice rotational symmetry. The wave functions constructed through this modified mapping should serve as better trial wave functions in the thermodynamical limit and on the rotationally invariant finite lattice.
%The wavefunctions constructed through this modified mapping should serve as better trial wavefunctions to compare with the numerics
Also these wave functions will form a good basis for the construction of lattice symmetry preserving pseudo-potential formalism for fractional Chern insulators. The focus of this paper shall be mainly on the $C_4$ rotational symmetry of square lattices. Similar analysis can be straightforwardly generalized to triangular or hexagonal lattices with $C_6$ symmetry. We also generalize the discussion to the
lattice symmetry of fractional Chern insulators with high Chern number bands.
\end{abstract}
\maketitle

\section{Introduction}

Integer and fractional quantum Hall states differ from the conventional phase of matter characterized via symmetry breaking by their distinct topological properties. Under strong magnetic field, the two-dimensional electron gas forms highly degenerate Landau levels (LLs) which carry non-trivial topological indices known as the TKNN invariants \cite{TKNN} or the Chern number. When a Landau level is fully occupied, the Chern number reveals itself as the quantized Hall conductance in an integer quantum Hall effect (IQH effect). When it comes to a partial filling, the strong correlation between electrons arise from the huge degeneracy of LL s(or the quenched kinetic energy), together with its non-trivial topology, leads to the fractional quantum Hall states (FQH states) that support fractionalized excitations and non-trivial braiding statistics. In a work in 1988\cite{HaldaneHoneycomb}, Haldane proposed the first realization of a band insulator model with non-trivial Chern bands (i.e.,  bands with non-zero Chern number). A fully occupied Chern band also shows a non-zero Hall conductance, which is referred to as a quantum anomalous Hall (QAH) insulator, or Chern insulator. This resemblance between the Chern bands and LLs is rooted in their equivalence in the sense of topology. Now it is conceivable that, if the Chern band is flat or almost flat, fractional Chern insulators (FCIs), which are FQH states without magnetic field, will emerge. Following this line of thought, lots of effort has been devoted into the search for FCI states\cite{XGFlatBand,KaiSunFlatBand,NSCMFlatBand,FQAH1,FQAH2,FQAH3,FQAH4,FQAH5,FQAH6,FQAH7,parameswaran2012,lee2012,Qi,barkeshli2012,sterdyniak2012,wang2012,liu2012,Lu2012}.
%XLnote: I removed reference to Haldane and TKNN here. I added some refs on higher chern number.

For the FQH states, wave functions, the validity of which can be checked most straightforwardly through numerics, have been a major tool to study the underlying physics. However, the understanding of FQH states through wave functions such as the Laughlin wave function is not directly applicable to FCI states, since the former can be written as holomorphic functions and the latter is defined on a lattice. As proposed in Ref. \onlinecite{Qi}, this difficulty can be overcome through the mapping between the LLs and Chern bands that exploit their similarity in the Wannier state representation (WSR). In a Chern band, the single particle Wannier states that are localized on one direction and carry well-defined wave vectors on the other resemble the Landau orbits in the LLs. By directly identifying them in the Fock basis, all FQH wave functions written in the continuum can be mapped into FCI wave functions. In the WSR, there is a gauge choice in the definition of the Wannier states in the Chern bands. Different gauge choices, though capturing the same topological properties, lead to microscopically different FCI states through the mapping. Attempts have been made to gauge-fix the Wannier states for the comparison with numerics
\cite{BernevigBlochModel2012,WuBernevigGaugeFix2012,Moller2012,liu2013}. Unfortunately, a generic gauge fixing scheme usually breaks the lattice rotation symmetry both in the thermodynamical limit and in the finite rotationally invariant lattice. In particular, for the gauge fixing scheme proposed in Ref. \onlinecite{WuBernevigGaugeFix2012}, the rotation symmetry is only restored when the Berry curvature in the Brillouin zone is homogeneous. Exceptionally, in Ref. \onlinecite{BernevigBlochModel2012}, an alternative gauge choice, i.e. the Coulomb gauge, is proposed. Although it is not constructed based on the crystal symmetry consideration, its Chern number $C=1$ version turns out to respect the four-fold lattice rotation symmetry. The symmetry property of the Coulomb gauge is not discussed in Ref. \onlinecite{BernevigBlochModel2012} and will be shown as a special case of our construction later. Although variational FCI wave functions that respects lattice rotation symmetry can be constructed through the projective construction \cite{Lu2012}, their connection to the FCI states exactly mapped from FQH states using WSR is still unclear.

In this paper, we focus on the construction of general gauging fixing schemes that respect the crystal rotation symmetry. Instead of choosing the gauge directly and verifying its properties under the crystal symmetry as the previous works did, we start with the requirement of the crystal rotation invariance of the FCI states proposed in Ref. \onlinecite{Qi} in Chern bands with $C=1$. Although this is a problem about many-body wave functions, it can be reduced to a single-particle one by noticing that if the singe-particle states in the Chern band transform in the same fashion under the crystal rotation symmetry as those in LLs, the rotation invariance of FCI states will be inherited from the its parent FQH states. We show that there is, in fact, a series of gauge choices which lead to this desired coincidence between the properties of the Chern bands states and LL states under the discrete symmetry on a single-particle level, including the one in Ref. \onlinecite{BernevigBlochModel2012} as one of the realizations. With these gauge choices, coherent states for the Chern bands and the pseudo-potential Hamiltonians that preserve the lattice rotation symmetry can also be constructed through this mapping between Chern number $C=1$ bands and LLs.\cite{Qi} Moreover, motivated by the construction of the lattice equivalence of multi-layer FQH in higher Chern number bands \cite{barkeshli2012} through a mapping that generalizes the one in Ref. \onlinecite{Qi}, we further generalize our discussion on symmetry preserving gauge fixing schemes to higher Chern number systems and, in particular, discuss the condition for the existence of such gauges for the $C=2$ case. The paper is organized as follows. First, we will discuss the general criteria for the gauges that respects the lattice rotation symmetry in the Chern bands with $C=1$. Secondly, we will use the lattice Dirac model \cite{QiWuZhang2006} and the checkerboard model \cite{KaiSunFlatBand} as examples to demonstrate the construction of the rotation symmetry preserving gauge for $C=1$ bands and explain the validness of this construction for Chern bands with $C=1$ in general. In the end, we will discuss the case with higher Chern numbers.

%while the construction in Ref. \onlinecite{Qi} is restricted to the Chern number $C=1$ bands, Ref. \onlinecite{barkeshli2012} proposes the construction of lattice equivalence of multi-layer FQH in Chern number bands through a similar mapping. Therefore,

%In this paper, we discuss the general procedure of constructing Wannier states that preserves crystal symmetry. Rather than using WSR, we equivalently identify the Bloch states of the LLs and Chern bands, similar to Ref. \onlinecite{BernevigBlochModel2012}. We show that there is a series of gauge choices which leads to Bloch states in Chern bands that transform in the same fashion as that in the LLs. With these gauge choices, coherent states for the Chern bands \cite{Qi} preserving the lattice rotation symmetry can be constructed. This implies that the FCI states and their corresponding pseudo-potential Hamiltonians constructed through the mapping between the Bloch states in the LLs and the Chern bands are invariant under the lattice rotation. In the following, we will first introduce the Bloch state representation of the LLs and discuss its rotation properties. Then we will use the lattice Dirac model \cite{QiWuZhang2006} and the Checkerboard model \cite{KaiSunFlatBand} as examples to demonstrate the construction of the rotation symmetry preserving gauge for the Chern bands with $C=1$. In the end, we will discuss the case with higher Chern number.

\section{General criteria}
%$\text{\it{Landau Levels}}$ -
In the original proposal \cite{Qi}, a LL is mapped to a Chern band by identifying the Landau orbits with the Wannier states. Equivalently, we can think of this mapping as an identification between Bloch states of the LL (that will be defined later) and those of the $C=1$ Chern bands.\cite{BernevigBlochModel2012} Although not explicit in this Bloch state representation, the LL does have a continuous $SO(2)$ rotation symmetry about the $z$ axis which includes a lattice rotation symmetry as a subgroup. Thus, lattice rotation symmetric FCI states can be obtained from $SO(2)$ rotation symmetric FQH states in the continuum written in the Fock basis by mapping the LL to the Chern bands in the Bloch state representation and requiring that the Bloch states in the LL and the Chern bands share the same transformation properties under the lattice rotation. As the first step, we should define the Bloch states of the LLs and study their behavior under lattice rotation. In the rest of the paper, we will mainly focus on the $C_4$ rotation symmetry of square lattices.

In the Landau gauge, the Landau orbits can be written as
\begin{align}
\psi_{p_{y}}(x,y)=\frac{1}{\sqrt{\sqrt{\pi} L_y l_b}}e^{ip_y y}e^{-\frac{1}{2l_b^2}(x-l_b^2p_y)^2},
\label{LandauGaugeWaveFunction}
\end{align}
where $L_y$ is the size of the system in the $y$ direction, $l_b=\sqrt{\hbar/(eB)}$ is the cyclotron length and $p_y=2\pi n/L_y$, $n\in\mathbb{Z}$ is the wavevector along the $y$ direction. As we can see, the Landau orbits are plane waves along the $y$ direction but localized along the $x$ direction. Now, we introduce a square lattice to the system with lattice constant $a$ satisfying $a^2=2\pi l_b^2$. Since each unit cell contains one flux quantum, the translations along the two base vectors commute with each other, which allows us to define Bloch states as an alternative basis of the LL. Consider an $N\times N$ system (in terms of lattice constant $a$). The Bloch wave function can be written as
\begin{align}
\psi^{\it bl}_{\bf k}(x,y)=\frac{1}{\sqrt{N}}\sum_{n\in\mathbb{Z}} e^{i k_x n a}\psi_{k_y+2n\pi/a }(x,y),
\end{align}
where ${\bf k}=(k_x,k_y)$ is the Bloch momentum with $k_x,k_y\in[-\frac{\pi}{a},\frac{\pi}{a}]$ and $k_{x,y}=2\pi n_{x,y}/(Na)$, $n_{x,y}\in \mathbb{Z}$.
Note that
\begin{align}
\psi^{\it bl}_{\bf k}=\psi^{\it bl}_{{\bf k}+{\bf g}_x}=e^{i k_x a}\psi^{\it bl}_{{\bf k}+{\bf g}_y}, \label{LLBZperiodicity}
\end{align}
where ${\bf g}_{x,y}$ are the two base vectors that indicate the periodicity of the Brillouin zone.
The Bloch wavefunction can be expressed in the standard form
\begin{align}
\psi^{\it bl}_{\bf k}(x,y)=e^{i k_x x+i k_y y} u_{\bf k}(x,y),
\end{align}
with
\begin{align}
&u_{\bf k} =\frac{1}{\sqrt{\sqrt{\pi}N^2 a l_b}}\sum_n  e^{i k_x (n a-x)+i \frac{2n\pi y}{a}}   e^{-\frac{(x-l_b^2k_y-n a)^2}{2l_b^2}}  \label{LLBloch}.
\end{align}
One can show straightforwardly that the Berry connection ${\bf A}$ in the Brillouin zone (BZ) is given by
\begin{align}
{\bf A}=-i\langle u_{\bf k}|\nabla_{\bf k}|u_{\bf k}\rangle=(-l_b^2 k_y,0),
\end{align}
from which we see that the Chern number of the LL is $C=1$ by integrating the Berry curvature across the whole BZ. From the definition of $u_{\bf k}(x,y)$ in Eq. \ref{LLBloch}, one can derive that
\begin{align}
u_{\bf k}( y, -x)=e^{-i k_x k_y l_b^2}e^{-i x y / l_b^2}u_{R{\bf k}}( x, y),  \label{LLBlochRotated}
\end{align}
where $R {\bf k}=(-k_y, k_x)$ is the Bloch momentum $\bf k$ rotated by $90^{\circ}$. Eq. \ref{LLBlochRotated} describes how the Bloch wavefunctions transform under the $C_4$ rotation in real space. On the right hand side of Eq. \ref{LLBlochRotated}, $u_{R{\bf k}}(x,y)$ is what would be normally expected for Bloch states under $C_4$ rotation. The factor $e^{-i x y / l_b^2}$ generates a real space gauge transformation which follows a spatial $C_4$ rotation and restores the apparently broken rotation symmetry by the Landau gauge. The factor $e^{-i k_x k_y l_b^2}$ results from a specific gauge choice of the Bloch states in the BZ of the LL.

To extract the gauge condition of the Bloch states that is crucial for $C_4$ rotation symmetry of the LLs, we will introduce the coherent states $\psi_{z_0}$ of the LL which form an overcomplete basis:
\begin{align}
\psi_{z_0}(x,y)=\frac{1}{\sqrt{2\pi }l_b} e^{i\frac{xy}{2l_b^2}} e^{i\frac{x_0y_0}{2l_b^2}} e^{-\frac{|z-z_0|^2+(\bar{z}z_0-z\bar{z}_0)}{4l_b^2}},
\end{align}
where $z_0=x_0+iy_0$ is the center of mass of the coherent states. These states are localized in both directions. And they are invariant under the $SO(2)$ spatial rotation around its center up to a real space gauge transformation. We can re-express them in terms of the Bloch states:
\begin{align}
\psi_{z_0}\propto \sum_n \sum_{k_{x,y}\in[-\frac{\pi}{a},\frac{\pi}{a}]}
e^{-\frac{y_0^2}{2l_b^2}} e^{-\frac{l_b^2}{2}(k_y+\frac{2n\pi}{a}-\frac{\bar{z}_0}{l_b^2})^2}e^{-i k_x n a}\psi^{\it bl}_{\bf k}.  \label{LLCoherent}
\end{align}
This expression will serve as the general definition for coherent states on Chern bands when they are mapped to the LLs. We notice that it is the gauge condition Eq. \ref{LLBZperiodicity} that guarantees the localization of $\psi_{z_0}$.
One also can verify that, under the $C_4$ rotation, the coefficients in the second line conspire with the factor $e^{-i k_x k_y l_b^2}$ to guarantee the invariance of $\psi_{z_0}$ under $C_4$.

Thus, if we want to establish an identification between the Bloch states of the LL and those of the Chern bands without breaking the lattice rotation symmetry, the Bloch states of the Chern bands should satisfy the same gauge condition: (1) Bloch states has the same periodicity as that in Eq. \ref{LLBZperiodicity}. (2) Under $C_4$ rotation, the Bloch wavefunction $u_{\bf k}$ satisfies:
\begin{align}
& u_{\bf k}=u_{{\bf k}+{\bf g}_x}=e^{i k_x a}u_{{\bf k}+{\bf g}_y}, \label{LLBZperiodicity2} \\
& u_{R{\bf k}}= \hat{R}_{C_4} e^{i k_x k_y l_b^2}u_{\bf k},   \label{LLGaugeRotation}
\end{align}
where $\hat{R}_{C_4}$ denotes the symmetry transformation inside the unit cell, which is $(x,y)\rightarrow (-y,x)$ followed by a real space gauge transformation in the LL case. As a convention, we can always require that $\hat{R}_{C_4}^4=1$ and $u_{{\bf k}=0}$ is the eigenstate of $\hat{R}_{C_4}$ with eigenvalue $1$.

\section{Examples}

$\text{\it{Lattice Dirac model}}$ - Now we use the lattice Dirac model\label{qi2006} as the first example to demonstrate how to preserve the $C_4$ symmetry by a gauge choice. The lattice Dirac model is given by the Hamiltonian $\hat{H}=\sum_{\bf k}c_{\bf k}^\dag \mathcal {H}({\bf k})c_{\bf k}$, where $c_{\bf k}$ is a two-component fermion operator and
\begin{align}
\mathcal {H}({\bf k})=\vec{\sigma}\cdot {\bf d}_ {\bf k},
\end{align}
with
\begin{align}
{\bf d}_{\bf k}=(\sin k_x,\sin k_y,M+\cos k_x +\cos k_y).
\end{align}
We have taken the lattice constant to be $a=1$. $M$ is a free parameter which is taken to be $0<M<2$ in order that the lower band has $C=1$. This model is symmetric under a simultaneous $C_4$ rotation in both the real space and spin space:
\begin{align}
\mathcal {H}(-k_y,k_x)=\hat{R}_{C_4} \mathcal {H}(k_x,k_y) \hat{R}^\dag_{C_4},
\end{align}
where $\hat{R}_{C_4}=e^{-i \frac{\pi}{4}}\left(\begin{array}{cc} e^{-i \frac{\pi}{4}} & 0 \\ 0& e^{i \frac{\pi}{4}}\end{array}\right)$ is the spin rotation matrix times a $U(1)$ charge rotation which is choosen, for later convenience, to satisfy $R_{C_4}^4=1$.

\begin{figure}[tb]
\centerline{
\includegraphics[width=0.45\textwidth]{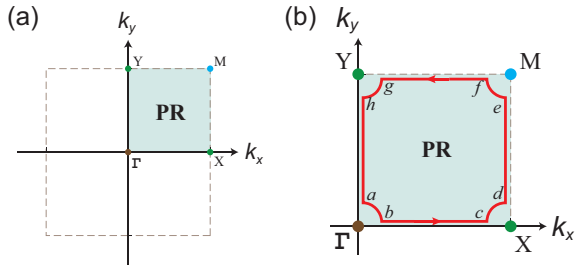}
}
\caption{(a) The BZ of square lattice with $k_{x,y}\in[-\pi,\pi]$ is indicated by the dashed line. The blue area is the PR with its four corners being high symmetry points: $\Gamma$, $X$, $Y$ and $M$. (b) The red line is a contour that follows the boundary of the PR. $\wideparen{ab}$, $\wideparen{cd}$, $\wideparen{ef}$ and $\wideparen{gh}$ are quarter circles with infinitesimal radii that centered around the four corners of the PR.
\label{PR}
}
\end{figure}

To gauge fix this Chern band, we can take the following two-step strategy. First, we write down the Bloch states in a gauge that respects the full periodicity of BZ and the $C_4$ rotation symmetry. Usually, this gauge will have singularities at the high symmetry points of the BZ. For the second step, we modify the gauge to satisfy the gauge condition Eq. \ref{LLBZperiodicity2} and Eq. \ref{LLGaugeRotation}. Since Eq. \ref{LLGaugeRotation} only connects the Bloch states with momentums that are related to each other by $C_4$ rotation, we can pick any gauge in the ``principle region" (PR) which is defined as a quarter of the BZ, $k_{x,y}\in[0,\pi/a]$ as shown in Fig. \ref{PR} (a), in this case. Then we use Eq. \ref{LLGaugeRotation} to generate the gauge on other parts of the BZ. The requirement of self-consistency and the consistency with gauge condition Eq. \ref{LLBZperiodicity2} imposes only a boundary condition to the the gauge choice in the PR.

Now, for the lattice Dirac model, we first write down the Bloch states in the lower band:
\begin{align}
|k_x,k_y\rangle=\left(\begin{array}{c} \sin\frac{\theta_{\bf k}}{2}e^{-i\varphi_{\bf k}}\\ -\cos\frac{\theta_{\bf k}}{2}\end{array}\right),  \label{LatticeDiracGauge1}
\end{align}
where $\cos\theta_{\bf k}=d^z_{\bf k}/|{\bf d}_{\bf k}|$ and $\varphi_{\bf k}=\arg (d^x_{\bf k}+i d^y_{\bf k})$. This gauge satisfies
\begin{align}
|k_x,k_y\rangle=|k_x+2\pi &,k_y\rangle=|k_x,k_y+2\pi\rangle, \nonumber \\
|-k_y,k_x\rangle & =\hat{R}_{C_4}|k_x,k_y\rangle.       \label{DLMGaugeS1}
\end{align}
It is easily shown that $|k_x,k_y\rangle$ is singular only at ${\bf k}=(\pi,\pi)$.
For step two, we denote the Bloch states in the Dirac model which maps to those of the LL with the $C_4$ symmetry preserved as $|k_x,k_y\rangle\rangle=e^{i \phi(k_x,k_y)}|k_x,k_y\rangle$. The gauge conditions Eq. \ref{LLBZperiodicity2} and Eq. \ref{LLGaugeRotation} then requires that
\begin{align}
e^{i \phi(-k_y,k_x)}=e^{i \frac{k_x k_y}{2\pi}} e^{i \phi(k_x,k_y)}, \nonumber \\
e^{i \phi(k_x,k_y)}=e^{i \phi(k_x+2\pi,k_y)}.\label{DLMGauge}
\end{align}
Notice that the "twisted" periodicity gauge condition along the $k_y$ direction in Eq. \ref{LLBZperiodicity2} can be generated by these two equations. Therefore, the phase factor $e^{i \phi(k_x,k_y)}$ in the BZ is completely defined by its value in the PR with the following consistency condition:
\begin{align}
\phi(k,0) &= \phi(0,k), \nonumber \\
\phi(k,\pi)+\frac{k}{2} &= \phi(-\pi,k)=\phi(\pi,k),  \label{DLMGaugeCondition1}
\end{align}
where $k\in[0,\pi]$.
Notice that the second equation leads to a singularity of $\phi({\bf k})$ on the boundary of the PR, since
\begin{align}
\lim_{k\rightarrow\pi}\phi(k,\pi)+\pi/2=\lim_{k\rightarrow\pi}\phi(\pi,k). \label{DLMGaugeCondition2}
\end{align}
This singularity correctly cancels that of $|k_x, k_y \rangle$ at $\bf{k}_M \equiv(\pi,\pi)$.
All functions $\phi({\bf k})$ in PR that satisfy the gauge condition Eq. \ref{DLMGaugeCondition1} and that are smooth, only except at $k_M$, can be consistently extended to the whole BZ by Eq. \ref{DLMGauge}. As an example, one solution is given by:
\begin{align}
\phi({\bf k})=
\left\{
\begin{array}{cc} \left(1-\frac{|{\bf k}-{\bf k}_M|}{\pi}\right)\theta'_{\bf k}, & \text{ for  } \frac{|{\bf k}-{\bf k}_M|}{\pi}<1\\ 0, & \text { elsewrhere in PR}
\end{array}\right. ,
\end{align}
where $\theta'_{\bf k}=\arg(k_x+i k_y-(1+i)\pi)-\pi$. %Also, one can follow the procedures proposed in \cite{BernevigBlochModel2012} to obtain a different special solution. %XLnote: This discussion will be made later so we probably shouldn't discuss it here since it will be difficult to understand.
Now we have obtained $|k_x,k_y\rangle\rangle$ which corresponds to $u_{\bf k}$ in the LL. With this identification, coherent states for the Chern band in the lattice Dirac model can be defined through Eq. \ref{LLCoherent}, and the $C_4$ rotation properties will be inherited from the LL.

At this stage, it seems that the cancellation of singularities between $\phi({\bf k})$ and $|k_x,k_y\rangle$ is quite accidental. However, we will show that it is, in fact, guaranteed by the rotation symmetry and the Chern number. In the first step of the construction, we always start with a gauge that satisfies Eq. \ref{DLMGaugeS1}. Such a gauge usually has singularities at the high symmetry points under rotation: the $\Gamma$ point with ${\bf k}_{\Gamma}=(0,0)$, the $X$ point with ${\bf k}_{X}=(\pi,0)$, the $Y$ point with ${\bf k}_{Y}=(0,\pi)$, and the $M$ point with ${\bf k}_{M}=(\pi,\pi)$. The singularities at these points are the $U(1)$ monodromies which can be determined by the eigenvalues of $\hat{R}_{C_4}$ at these points. At the $\Gamma$ point, by definition,
\begin{align}
\lim_{\epsilon\rightarrow 0}|{\bf k}_{\Gamma}+\epsilon {\bf g}_y\rangle  =\hat{R}_{C_4}|{\bf k}_{\Gamma}+\epsilon {\bf g}_x\rangle.
\end{align}
Meanwhile, $|{\bf k}_{\Gamma}\rangle$ must be an eigenstate of the $\hat{R}_{C_4}$ with its eigenvalue denoted as $\xi_{\Gamma}$. Thus, we have
\begin{align}
|{\bf k}_{\Gamma}+\epsilon {\bf g}_y\rangle \sim \xi_{\Gamma}|{\bf k}_{\Gamma}+\epsilon {\bf g}_x\rangle,~~~\epsilon\rightarrow0.  \label{GammaSing}
\end{align}
Thus, the $U(1)$ phase singularity along the infinitesimal quarter circle $\wideparen{ab}$ is $\xi_{\Gamma}^{-1}$. Parallel analysis shows that, at the $M$ point, the inverse of the eigenvalue of $\hat{R}_{C_4}$ at the $M$ point denoted as $\xi_M$ equals the $U(1)$ phase singularity along the path $\wideparen{ef}$. For the $X$ and $Y$ points, they are only invariant under $\hat{R}^2_{C_4}$. We denote the eigenvalue of $\hat{R}^2_{C_4}$ at $X$ as $\eta_X$ which equals that at the $Y$ point. Given the gauge choice that satisfies Eq. \ref{DLMGaugeS1}, through a similar reasoning, we also find that total $U(1)$ phase singularity along the path $\wideparen{cd}$ and $\wideparen{gh}$ is equal to $\eta_X^{-1}$. Now we have related the $U(1)$ phase singularities of the gauge to the eigenvalues of the $\hat{R}_{C_4}$ and $\hat{R}_{C_4}^2$ at these high symmetry points. And it is shown in Ref. \onlinecite{Bernevig2012SymChern} that these eigenvalues are connected to the Chern number of the band through the following formula:
\begin{align}
e^{i \frac{2\pi C}{4}}=\xi_{\Gamma}^{-1}\xi_{M}^{-1}\eta_X^{-1}. \label{ChernSym}
\end{align}
Thus, for $C=1$, the total amount of $U(1)$ singularity of the gauge $|k_x,k_y\rangle$ at four high symmetry points is given by $e^{i \frac{2\pi }{4}}$ which matches the singularity of the phase factor $e^{i\phi({\bf k})}$ required by Eq. \ref{DLMGaugeCondition1}. As a result, we have shown that the cancellation of singularity of the gauge and that of the phase factor $e^{i\phi({\bf k})}$ is guaranteed.

Here we have used the result Eq. \ref{ChernSym} from Ref. \onlinecite{Bernevig2012SymChern}. This is an important step in our justification of the cancellation between singularities. In the current setting, we can provide a simple explanation of Eq. \ref{ChernSym} which helps to understand our justification. At the $\Gamma$ point, given the gauge choice with the singularity shown in Eq. \ref{GammaSing}, it is straightforward to find that the exponentiated Berry connection integral along the path $\wideparen{ab}$ equals $\xi_{\Gamma}^{-1}$. Similarly, the exponentiated Berry connection integral along the path $\wideparen{ef}$ equals $\xi_M^{-1}$ at the $M$ point. And along $\wideparen{cd}$ and $\wideparen{gh}$, the exponentiated total Berry connection integral should be equal to $\eta_X^{-1}$. Also, we notice that, due to the gauge condition Eq. \ref{DLMGaugeS1}, the Berry connection integral along $\overline{ha}$ cancels that along $\overline{bc}$. The same happens for $\overline{de}$ and $\overline{fg}$. Hence, the exponentiated total Berry connection integral
along the red contour in Fig. \ref{PR} only receives contribution from the four infinitesimal quarter circles ($\wideparen{ab}$, $\wideparen{cd}$, $\wideparen{ef}$, and $\wideparen{gh}$) and is given by $\xi_{\Gamma}^{-1}\xi_{M}^{-1}\eta_X^{-1}$. On the other hand, this exponentiated total Berry connection integral
along the red contour should equal $e^{i \frac{2\pi C}{4}}$ because the red contour encloses the PR which contains $1/4$ of the total Berry curvature in the Brillouin zone due to the $C_4$ rotation symmetry. Now we have rederived Eq. \ref{ChernSym}.

\begin{figure}[tb]
\centerline{
\includegraphics[width=0.45\textwidth]{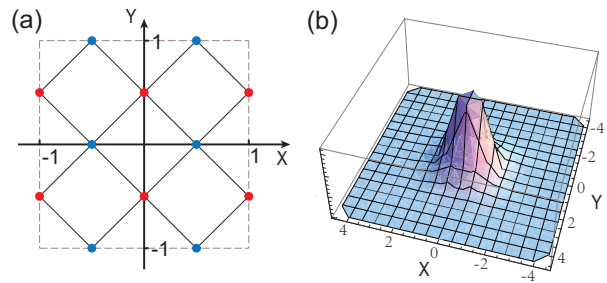}
}
\caption{(a) On the checkerboard lattice, the red and blue sites are the upper and lower components of the Bloch state. The coordinates are given in the unit of lattice constant. (b) The wavefunction of a coherent state that centers at the origin is symmetric under $C_4$ rotation.
\label{Checkerboard}
}
\end{figure}

$\text{\it{Checkerboard model}}$ - As a second example system where $C_4$ symmetry acts differently, we consider the checkerboard model that produces a flat Chern band \cite{KaiSunFlatBand}:
\begin{align}
\mathcal {H}({\bf k})=\epsilon_{\bf k} I+h^x_{\bf k} \sigma_x+h^y_{\bf k} \sigma_y+h^z_{\bf k} \sigma_z,  \label{KaiSunHam}
\end{align}
where
\begin{align}
{\bf h}_{\bf k}=(\cos \frac{k_x}{2}\cos \frac{k_y}{2}, -\sin \frac{k_x}{2}\sin \frac{k_y}{2},\frac{\cos k_x-\cos k_y}{2\sqrt{2}+2})
\end{align}
and $\epsilon_{\bf k}=\frac{1}{\sqrt{2}+2}\cos k_x \cos k_y$ that does not affect the Bloch states. The BZ is defined as $k_x,k_y\in[-\pi,\pi]$ with the gauge condition
\begin{align}
\sigma_z\mathcal {H}(k_x,k_y)\sigma_z= \mathcal {H}(k_x+2\pi,k_y) = \mathcal {H}(k_x,k_y+2\pi).
\end{align}
The $C_4$ symmetry and gauge property of this model is given by
\begin{align}
\sigma_x\mathcal {H}(-k_y,k_x)\sigma_x=& \mathcal {H}(k_x,k_y).
\end{align}
With the reference of the gauge condition Eq. \ref{LLBZperiodicity2} and Eq. \ref{LLGaugeRotation}, we need to find the Bloch states in a gauge such that:
\begin{align}
|k_x+2\pi,k_y\rangle\rangle &=   \sigma_z|k_x,k_y\rangle\rangle  \nonumber \\
|-k_y,k_x\rangle\rangle &= -\sigma_x e^{i\frac{k_x k_y}{2}} |k_x,k_y\rangle\rangle.  \label{KaiSunSymCondition}
\end{align}
Following the two-step analysis described in the lattice Dirac model case, we obtain the general solution:
\begin{align}
|k_x,k_y\rangle\rangle= e^{i\phi({\bf k})} \left(\begin{array}{c} -\sin\frac{\theta_{\bf k}}{2}\\ \cos\frac{\theta_{\bf k}}{2}e^{i\varphi_{\bf k}}\end{array}\right),  \label{KaiSunSolution}
\end{align}
where $\cos\theta_{\bf k}=h^z_{\bf k}/|{\bf h}_{\bf k}|$, $\varphi_{\bf k}=\arg (h^x_{\bf k}+i h^y_{\bf k})$ and $e^{i\phi({\bf k})}$ is a $U(1)$ phase that satisfies the following condition only on the boundary of the PR:
\begin{align}
\phi(k,0)&=\phi(0,k) \nonumber \\
\phi(\pi,k)&=\frac{k}{2}-\frac{\pi}{2}+\phi(k,\pi),
\end{align}
where $k\in[0,\pi]$.
As expected, the singularity of the $\phi({\bf k})$ around the boundary of PR can be set to cancel the singularity of the two-component spinor in Eq. \ref{KaiSunSolution}. Now $C_4$ symmetric coherent states can be constructed via replacing $\psi^{\it bl}_{\bf k}$ in Eq. \ref{LLCoherent} with the Bloch states in Eq. \ref{KaiSunSolution}. The wave function of a coherent state that centers at the origin is depicted in Fig. \ref{Checkerboard} (b).

At the end of this section, let us discuss why the Coulomb gauge in Ref. \onlinecite{BernevigBlochModel2012} preserves the $C_4$ symmetry. In the Coulomb gauge, the gauge potential ${\bf A}_{\bf k}$ is the sum of two parts ${\bf A}={\bf A}_0+\delta {\bf A}$ with ${\bf A}_0$ identical to the gauge potential in the LL, and the correction $\delta {\bf A}$ with zero Chern number. Once gauge fixed by the Coulomb gauge condition, $\delta{\bf A}$ is single-valued in the BZ and transforms trivially under $C_4$ rotation. Therefore, the $C_4$ rotation property of the Coulomb gauge is completely determined by ${\bf A}_0$ and thus is identical to the LL. Therefore the Coulomb gauge preserves the $C_4$ rotation symmetry in the mapping to the LL.

 \section{Generalization to higher Chern number systems}

%$\text{\it{Models with higher Chern number}}$ -
So far we have demonstrated that, for $C=1$ Chern bands, we can always find the appropriate gauge such that the $C_4$ symmetry is preserved when the Chern band is mapped to the LL. Now we generalize the discussion to the case of Chern bands with Chern number $C>1$. It has been shown in Ref. \onlinecite{barkeshli2012} that a Chern band with Chern number $C$ can be mapped into $C$ layers of Chern number $1$ bands by WSR (given the condition that the system size is commensurate with $C$ along the direction in which the Wannier states are localized). An alternative approach has been developed in Ref. \onlinecite{BernevigBlochModel2012}. Since this mapping generically breaks the rotation symmetry, the existence of a rotationally symmetric scheme requires further inspection. In the following we will focus on the $C_4$ symmetry of FCI's in a $C=2$ Chern band.

As is discussed in Ref. \onlinecite{barkeshli2012}, in general, one-dimensional Wannier states of the $C=2$ band can be defined by one-dimensional Fourier transform of the Bloch states $\left|{{\bf k}}\right\rangle$ along a reciprocal vector direction $n\hat{x}+m\hat{y}$. For each pair of mutually prime integers $(n,m)$, a set of Wannier states are defined, which defines a one-to-one mapping between bilayer FQH states and the FCI in this $C=2$ band. If $n$ is odd, the resulting FCI states are sensitive to the translation along $x$ direction, such that an $x$-dislocation with Burgers vector ${\bf b}=\hat{x}$ will generically become a non-Abelian defect. Similarly, if $m$ is odd the $y$-dislocation with Burgers vector ${\bf b}=\hat{y}$ becomes a non-Abelian defect. Therefore the topological properties of the FCI state necessarily break $C_4$ symmetry unless $n$ and $m$ are both odd. In the following we will study the $C_4$ symmetry of the simplest case $(n,m)=(1,1)$ and discuss the more general cases later.

The Wannier states for $(n,m)=(1,1)$ are defined as
\begin{eqnarray}
\ket{W_n(k_y)}=\frac1{\sqrt{L}}\sum_{k_+\in[0,2\pi)}e^{ik_+n}\ket{k_+,k_++k_y}\label{WC2}
\end{eqnarray}
with $L=L_x=L_y$ the linear size of the system along both directions, and $\ket{k_x,k_y}$ the Bloch state of the $C=2$ band. $k_+$ denotes the momentum along the diagonal direction and $k_y$ denotes the momentum along the $y$ direction at $k_+=0$. The Bloch state can have different gauge choices and the goal of the following discussion is to find the conditions the $C_4$ symmetry impose to the gauge choice. For $C=2$  bands, the Wannier states at even and odd sites form two families which are not connected by adiabatically changing $k_y$. Each family is topologically equivalent to a $C=1$ band, or a Landau level. Since the FCI state constructed by WSR are bilayer FQH states with each family mapped to a layer of Landau level, the $C_4$ symmetry will be preserved if it is preserved for each layer. The Bloch states of each family can be obtained by an inverse Fourier transformation to each family of Wannier states in Eq. \ref{WC2}:
\begin{eqnarray}
\ket{k_+,k_++k_y}_1&=&\frac1{\sqrt{L/2}}\sum_{n}\ket{W_{2n-1,k_y}}e^{-2ink_+}\nonumber\\
\ket{k_+,k_++k_y}_2&=&\frac1{\sqrt{L/2}}\sum_{n}\ket{W_{2n,k_y}}e^{-2ink_+}.\label{reducedBloch}
\end{eqnarray}
Here $k_+\in[0,\pi)$ and $k_y=[0,2\pi)$ defines the reduced Brillouin zone. Combining Eq. \ref{WC2} and  \ref{reducedBloch}, we obtain the following simple relation between the Bloch states of the $C=2$ states and that of the effective $C=1$ bands:
\begin{eqnarray}
\ket{k_+,k_++k_y}_1&=&\frac{e^{-ik_+}}{\sqrt{2}}\left(\ket{k_+,k_++k_y}-\ket{k_++\pi,k_++k_y+\pi}\right)\nonumber\\
\ket{k_+,k_++k_y}_2&=&\frac1{\sqrt{2}}\left(\ket{k_+,k_++k_y}+\ket{k_++\pi,k_++k_y+\pi}\right)\nonumber
\end{eqnarray}
or equivalently, in the conventional $k_x,k_y$ basis,
\begin{eqnarray}
\ket{{\bf k}}_1&=&\frac1{\sqrt{2}}\left(\ket{{\bf k}}-\ket{{\bf k}+(\pi,\pi)}\right)e^{-ik_x}\nonumber\\
\ket{{\bf k}}_2&=&\frac1{\sqrt{2}}\left(\ket{{\bf k}}+\ket{{\bf k}+(\pi,\pi)}\right).\label{eq:folding}
\end{eqnarray}
To preserve $C_4$ symmetry in the two $C=1$ bands, the Bloch states $\ket{{\bf k}}_{1,2}$ need to satisfy the $C_4$ invariance conditions (\ref{LLBZperiodicity2}) and (\ref{LLGaugeRotation}). Eq. \ref{eq:folding} together with Eq. \ref{LLBZperiodicity2} and \ref{LLGaugeRotation} determines the general $C_4$ symmetry condition for the $C=2$ FCI states constructed in the $(1,1)$ WSR.

For more generic $(n,m)$ with $n,m$ odd, the discussion is exactly parallel and the reduced bands $\ket{{\bf k}}_{1,2}$ are determined by Eq. \ref{eq:folding} with the momentum $Q=(\pi,\pi)$ replaced by $Q'=(n\pi,m\pi)$, and the factor $e^{-ik_x}$ replaced by $e^{-i(t k_x-s k_y)}$ with $s,t$ two integers satisfying $\det\left(\begin{array}{cc} n & m \\ s & t \end{array}\right)=1$.
%$k_1$ the momentum component along the $(n,m)$ direction.
For $n,m$ odd, $Q'$ and $Q$ are equal modulo a reciprocal lattice vector. Therefore the effective $C=1$ bands for all $(n,m)$ odd are equivalent up to a phase factor.

\begin{figure}[tb]
\centerline{
\includegraphics[width=0.28\textwidth]{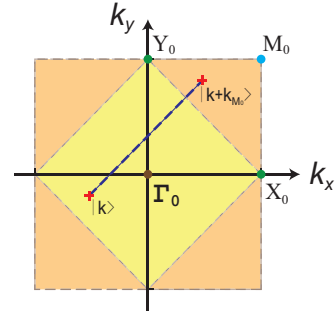}
}
\caption{A Chern number $C=2$ band is mapped into two subbands with $C=1$ by folding the BZ, namely taking a superposition of the states $|{\bf k}\rangle$ with $|{\bf k}+{\bf k}_{M_0 }\rangle$.
\label{folding}
}
\end{figure}

In particular, the $C_4$ invariance condition imposes some constraints on the rotation eigenvalues of the $C=2$ band at the high symmetry points. The original and reduced BZ's are shown in Fig. \ref{folding}. In the same way as the $C=1$ band discussed in the last section, we denote the $R_{C_4}$ eigenvalue of the original $C=2$ band at high symmetry points $\Gamma_0,M_0$ as $\xi_{\Gamma_0},\xi_{M_0}$, and the $R_{C_4}^2$ eigenvalue at $X_0$ point as $\eta_{X_0}$. At the $\Gamma$ points of the reduced BZ, the $C=1$ Bloch states $\ket{{\bf k}}_{1,2}$ are superpositions of $\ket{(0,0)}$ and $\ket{(\pi,\pi)}$ in the original BZ.  Therefore the $C=1$ bands at $\Gamma$ point can only be $C_4$ invariant if $\xi_{\Gamma_0}=\xi_{M_0}$. The $\hat{R}_{C_4}$ eigenvalue of the two $C=1$ bands at the $\Gamma$ point satisfies $\xi_{\Gamma_{1,2}}=\xi_{\Gamma_0}=\xi_{M_0}$. At the $M$ points of the subbands, the Bloch states are the superpositions of $|{\bf k}_{X_0}\rangle_0$ and $|{\bf k}_{Y_0}\rangle_0$ which are both eigenstates of $\hat{R}_{C_4}^2$ with eigenvalue $\eta_{X_0}$. Hence, the eigenvalue of $\hat{R}_{C_4}$ at $M_{1,2}$ should satisfy $\xi_{M_{1,2}}^2={\eta_{X_0}}$. %The $\pm$ sign for each of $\xi_{M_{1,2}}$ cannot be determined in this analysis.
At the $X$ points of the subbands, the Bloch states are the superposition of $|{\bf k}=(-\frac{\pi}{2},-\frac{\pi}{2})\rangle$ and $|{\bf k}=(\frac{\pi}{2},\frac{\pi}{2})\rangle$. Since $\hat{R}_{C_4}^4=1$ by definition, the eigenvalue of $\hat{R}_{C_4}^2$ at $X_{1,2}$ should be $\eta_{X_{1,2}}^2=1$. %Similarly, the sign of each of $X_{1,2}$ cannot be determined here.
By construction the two bands $\ket{{\bf k}}_{1,2}$ each have Chern number $C=1$, so that %From the assumption that the original band has $C=2$ and both subbands have $C=1$,
we obtain based on Eq. \ref{ChernSym} that $\xi^{-1}_{\Gamma_0}\xi^{-1}_{M_0}\eta^{-1}_{X_0}=-1$ and $\xi^{-1}_{\Gamma_j}\xi^{-1}_{M_j}\eta^{-1}_{X_j}=i$ with $j=1,2$.
To summarize, the $C_4$ symmetry requires the following conditions to the rotation eigenvalues of the original band and the effective $C=1$ bands:%for the splitting to be possible, we require (with a proper choice of $\pm$ signs):
\begin{align}
&\xi^{-1}_{\Gamma_0}\xi^{-1}_{M_0}\eta^{-1}_{X_0} = -1, \nonumber\\
&\xi^{-1}_{\Gamma_1}\xi^{-1}_{M_1}\eta^{-1}_{X_1}=\xi^{-1}_{\Gamma_2}\xi^{-1}_{M_2}\eta^{-1}_{X_2}=i, \nonumber \\
&\xi_{M_{1}}^2=\xi_{M_2}^2={\eta_{X_0}},~~\eta_{X_{1}}^2=\eta_{X_2}^2=1,\nonumber\\
&\xi_{\Gamma_{1}}=\xi_{\Gamma_2}=\xi_{\Gamma_0}=\xi_{M_0}.
\end{align}
Since $\xi_{\Gamma_0}=\xi_{M_0}$ is required, it indicates that there are some $C=2$ bands which can never support any FCI state with $C_4$ symmetry, at least not one constructed by WSR.
For example, consider the model:
\begin{align}
\mathcal {H}(k_x,k_y) = & (\sin^2k_x-\sin^2k_y) \sigma_x + 2\sin k_x\sin k_y \sigma_y  \nonumber \\
&+(1+\cos k_x+\cos k_y) \sigma_z
\end{align}
with $\hat{R}_{C_4}=\sigma_z$. The lower energy band has $\xi_{\Gamma_0}=-1,~\xi_{M_0}=1$, so that no $C_4$ invariant gauge choice can be found.

In summary, for a $C=2$ Chern band on the square lattice, a scheme of mapping it into two $C=1$ bands while preserving the $C_4$ rotation symmetry does NOT always exist. A possible obstruction can be detected by calculating the eigenvalue of the $C_4$ operation at high symmetry points in the BZ. The question remains: What are the conditions that guarantee a $C_4$ symmetry preserving splitting scheme of a Chern band with a generic Chern number $C$? It is very likely that a generic Chern number will require a enlargement of unit cell that is incompatible with the rotation symmetry and, thus, will forbid a symmetry preserving splitting of the Chern band. This may be a reason to believe that the topological nematic state is generic.

\section{Conclusion and discussions}
%$\text{\it{Conclusion and discussion}}$ -
In conclusion, we propose a scheme to select the gauge of the Bloch states of $C=1$ Chern bands so that the mapping from LL to the Chern bands respects the $C_4$ rotation symmetry of the square lattice. Consequently, FCI states obtained through this mapping are invariant under the $C_4$ rotation, and thus, are better model wave functions in the thermodynamical limit.
Also on rotationally invariant finite lattice, they should have better overlap with numerical simulation due to the same symmetry reason. And even for a generic $N_1\times N_2$ lattice that explicitly breaks the rotation symmetry, as long as the correlation length is much smaller than lattice size, the correction of the FCI states due to the non-unit aspect ratio of the finite should be small. In this case, the trial wave functions we proposed should still be good candidates to compare with the numerics. Also, through the $C_4$ invariant coherent states, a $C_4$ symmetric pseudo-potential formalism can be constructed. Our discussion mainly focuses on the $C=1$ Chern bands with $C_4$ symmetry on square lattices. Similar construction can be done for those with $C_6$ symmetry on triangular or hexagonal lattices. The discussion is generalized to a higher Chern number case.
%We provide examples in which the $C=2$ bands can be split into two $C=1$ bands without breaking the $C_4$ rotation symmetry. We also identify constraints to the splitting in a generic model with $C=2$ and the $C_4$ symmetry. However,
A $C=2$ band can be reduced to two $C=1$ bands by WSR. The $C_4$ symmetry property of the $C=2$ band can be determined by that of the two reduced $C=1$ bands. $C_4$ symmetry imposes some constraints on the direction of the Wannier states in WSR, and also some constraints on the rotation eigenvalues at the high symmetry points. The conditions that guarantee a rotationally symmetric splitting scheme for a Chern band with an arbitrary Chern number $C$ needs further investigation. Moreover, it is worthy to point out that so far our discussion has been focusing on the exact mapping between higher Chern number bands and multi-layer LLs through which  crystal rotationally invariant FCI states can be constructed from $SO(2)$ rotationally invariant multi-layer FQH states. However, as proposed in Ref. \onlinecite{Lu2012}, variational FCI states that preserve the crystal symmetry can always be constructed through projective construction for bands with any Chern number. The connection between these variational FCI states and the ones constructed from the mapping through WSR remains to be studied.

$\text{\it{Acknowledgements}}$ - This work is supported by the David and Lucile Packard Foundation.

\bibliography{TI}

\end{document}